\documentclass[aps,prl,amsmath,amssymb,reprint,superscriptaddress,floatfix]{revtex4-2}

\usepackage{lipsum}
\usepackage{bm}
\usepackage[normalem]{ulem}
\usepackage[utf8]{inputenc}
\usepackage[T1]{fontenc}
\usepackage{hyperref}
\usepackage[capitalize]{cleveref}
\usepackage{appendix}
\usepackage{cancel}
\usepackage{todonotes}
\usepackage{physics}
\usepackage{booktabs}
\usepackage{siunitx}
\usepackage[caption=false]{subfig}
\usepackage[bbgreekl]{mathbbol}
\usepackage{color,soul}
\usepackage{graphicx}
\graphicspath{{figures/}}
\usepackage{threeparttable}
\usepackage{comment}

\newcommand{\rp}{r_\parallel}

\newcommand{\ct}{c_\mathrm{T}}
\newcommand{\ctz}{c_\mathrm{T0}}

\newcommand{\cto}{c_\mathrm{T1}}
\newcommand{\cl}{c_\mathrm{L}}
\newcommand{\vecu}{\boldsymbol{u}}

\newcommand{\vrp}{\bm{\rp}}

\newcommand{\zela}{\zeta^{\mathrm{ela}}}
\newcommand{\zvis}{\zeta^{\mathrm{vis}}}

{\end{aligned}
\end{equation}}

\newcommand{\na}{n_\mathrm{A}}
\newcommand{\REMOVE}[1]{}

\newcommand{\neta}{\tilde{\eta}}

\begin{document}

\preprint{APS/123-QED}
\title{Friction on layered media: How deep do phonons reach?}
\author{Miru Lee}
\email{miru.lee@uni-goettingen.de}
\affiliation{Institute for Theoretical Physics, Georg-August-Universit\"at 
G\"ottingen, 37073 G\"ottingen, Germany}
\author{Niklas Weber}
\affiliation{Institute of Materials Physics, Georg-August-Universit\"at 
G\"ottingen, 37073 G\"ottingen, Germany}
\author{Cynthia A. Volkert}
\affiliation{Institute of Materials Physics, Georg-August-Universit\"at 
G\"ottingen, 37073 G\"ottingen, Germany}
\author{Matthias Kr\"uger}
\email{matthias.kruger@uni-goettingen.de}
\affiliation{Institute for Theoretical Physics, Georg-August-Universit\"at 
G\"ottingen, 37073 G\"ottingen, Germany}

\date{\today}

\begin{abstract}
We theoretically study the frictional damping of a small probe object on a 
coated planar surface, analyzing the resulting phonon modes via a theory of 
viscoelasticity. Three different types of excitations are found to contribute 
to friction in distinct ways: traveling (3D) spherical waves, traveling (2D) 
surface waves, and evanescent waves. While traveling waves transport energy 
away from the probe, determined by long range elastic properties (wavelength), 
evanescent waves transform energy into heat in a near-field range, 
characterized by the size of the probe. Thus, fundamentally different behaviors 
are predicted, depending on coating thickness and material properties.
\end{abstract}
\maketitle
Sliding friction is a complex phenomenon, involving surface 
mechanics~\cite{prandtl28a,muser11a,persson13a,persson03a,gnecco00a,socoliuc04a,qu20a, wada18a}
 and production of heat in the surrounding media. In the latter process, the 
relevant degrees of freedom have been found to include  
electronic~\cite{boldin18a,volokitin01a,dorofeyev99a,volokitin07b,kisiel11a_experiment}
 and  phononic 
ones~\cite{volokitin07b,dorofeyev99a,stipe01a_experiment,gotsmann01a_experiment,persson85a,volokitin06a,hu20a,kisiel11a_experiment,weber21a,schmidt20a,kantorovich08a_theory,kantorovich08b_theory,panizon18a, wada18a}.
 The role of phonon modes can be isolated by tuning them while keeping the 
surface properties and  contact mechanics unchanged. This has been achieved by 
inducing a phase transition in the solid~\cite{weber21a,kisiel11a_experiment}, 
or by changing an external electric field~\cite{schmidt20a}.

Studying friction on layers of different thickness is another way of achieving 
this goal; it is especially insightful as it not only allows material 
properties to be tailored while limiting changes in the contact surface, it 
also reveals how deep friction feels into the material.
Many numerical and nanoscale experimental studies have been done on the effect 
of 
layers~\cite{daly96a,kajita09a,xu11a,smolyanitsky12a,smolyanitsky12b,smolyanitsky15a,filleter09a,filleter10a,lee10a_gp,berman14a,benassi10a_theory,deng12a,li16a},
 with a variety of interesting behaviors. For example, while adding graphene 
layers between sliding bodies can substantially reduce 
friction~\cite{berman14a}, both increases~\cite{daly96a,kajita09a,xu11a} and 
decreases~\cite{filleter09a,filleter10a,lee10a_gp,smolyanitsky12a} in friction 
have been observed as the number of graphene layers is increased. Friction has 
also been observed to decrease with thickness for other layered materials, such 
as molybdenum disulfide and niobium diselenide, when they are weakly anchored 
to the substrate~\cite{lee10a_gp}.
A numerical study, in contrast, found that friction increases with thickness, 
when the bottom layer absorbs incoming phonons~\cite{benassi10a_theory}. 
Additionally, it has been reported that friction is smaller for strongly 
anchored samples compared to weakly anchored ones, suggesting that the changes 
in friction could be results of changes in local 
deformations~\cite{smolyanitsky12b,smolyanitsky15a,li16a}.
The complexity of the observed behavior calls for a theoretical  analysis that 
systematically addresses the dependence of phononic damping on layer thickness, 
boundary conditions, and material properties.

In this manuscript, we analyze friction of a nanoscopic object on a 3D planar 
coated substrate, treating phonon modes via a field theory of viscoelasticity 
that includes phonon attenuation. We find that friction arises  due to 
traveling spherical waves, cylindrical surface waves, or evanescent waves, each 
dominating in different regimes of material properties and coating thickness. A finite viscosity thus not only results in phonon attenuation, but also, here more importantly, in losses from evanescent waves.
Consequently, friction shows drastically different dependencies on coating 
thickness ranging from short range to long range behavior, and can increase or 
decrease with coating thickness. These regimes are  determined by the phonon 
attenuation coefficients and the refractive index.

\begin{figure}
    \centering
    \includegraphics[width=0.35\textwidth]{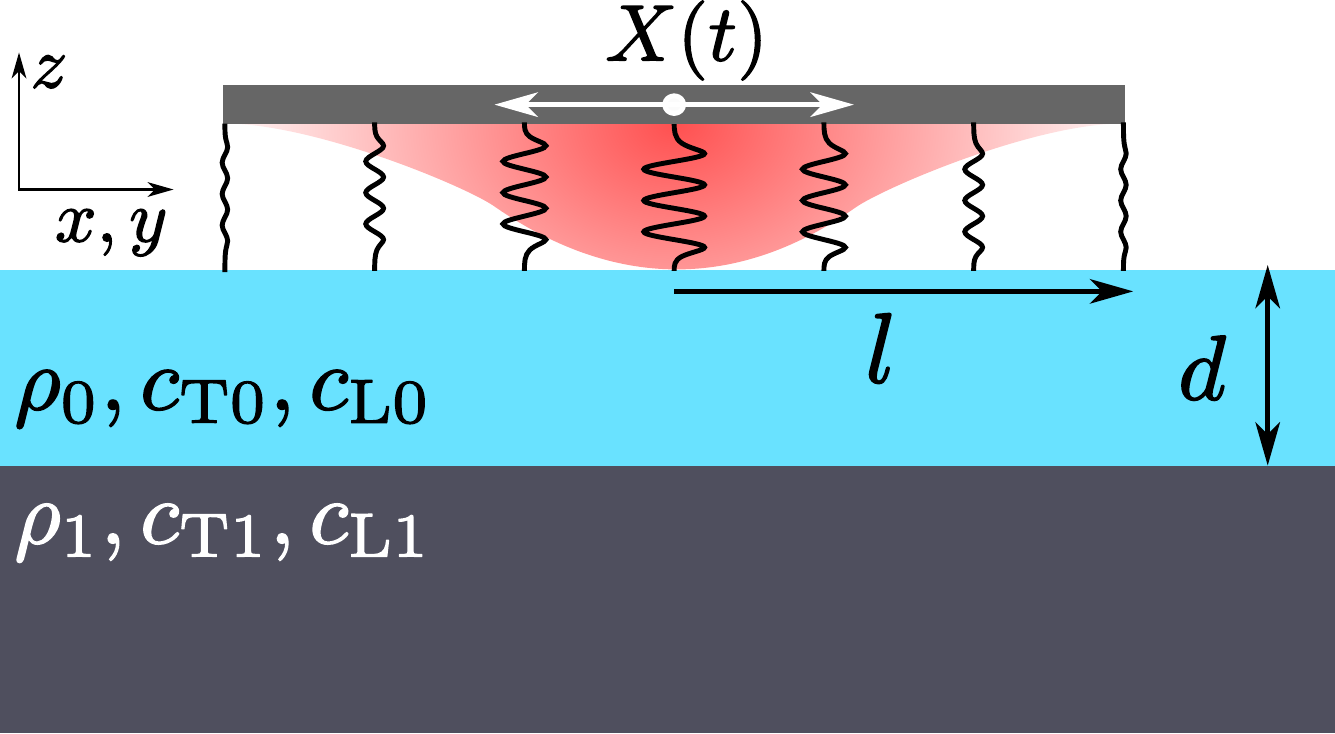}
    \caption{Investigated system: a probe is coupled to the surface of a coated 
    substrate, and oscillates parallel to it. The coupling interaction carries 
    an interaction radius $l$. The coating of thickness $d$ and the substrate 
    are characterized by mass density $\rho$ and transverse $(\ct)$ and 
    longitudinal $(\cl)$ speeds of sound, as indicated.  
}
    \label{fig:drawing}
\end{figure}

Consider a probe coupled to an isotropic solid filling the space $z\leq 0$ 
(see~\cref{fig:drawing}).
The probe oscillates parallel to the surface, so that its $x$ coordinate is 
$X(t)=\Re{ X_0 e^{i\omega t}}$.
The coupling with the surface causes a force acting on the probe, whose 
$x$-coordinate is $F(t)=\Re{F_0 e^{i\omega t}}$.
The damping coefficient or friction coefficient of the probe $\Gamma$ is defined as $ 
\Gamma=\omega^{-1}\Im{F_0/X_0}$ \cite{risken96a}

The specific form of coupling is not important for the conclusions of this 
manuscript, as detailed in the Supplemental Material (SM). We thus use a simple 
linear one, which allows us to find analytic expressions,
\begin{equation}
\begin{aligned}
    F(t)=&\na\kappa\iint_{-\infty}^{\infty}\dd[2]{\vrp}
    (-X(t)+u_x(\vrp,t))e^{-\frac{\vrp^2}{l^2}},
\label{eq:vfps}
\end{aligned}
\end{equation}
where $\vrp=(x,y,0)$ marks a position on the surface, and ${u}_x(\vrp,t)$ is 
the $x$-component of the phonon field at $z=0$ (introduced below). $\na$ is the 
particle number per unit area \cite{hamaker37a}, and $\kappa$ is the coupling 
strength.  \Cref{eq:vfps} contains a Gaussian envelope of width $l$, 
introducing a length scale of the interaction range or probe size.

The first term in~\cref{eq:vfps} is the force in absence of phonon
excitations; it is in phase with $X(t)$ and does not contribute to the damping 
coefficient $\Gamma$. The second term is the force due to excitations of the 
phonon field $\bm{u}$, treated via a theory of viscoelasticity, i.e., using a 
Kelvin-Voigt model~\cite{findley13a,landau86a,lee55a,lee21a},
\begin{equation}
\left(\cl^2-\ct^2\right)\nabla \nabla\cdot\vecu(\bm{r},\omega)+ \ct^2 
\nabla^2\vecu(\bm{r},\omega)=-\omega^2\vecu(\bm{r},\omega).
\label{eq:eqm}
\end{equation}
The coupling to the probe enters~\cref{eq:eqm} via a time dependent boundary 
condition. The resulting solution for $u_x$  
contains a part that is phase shifted with respect to $X(t)$, yielding the 
damping coefficient $\Gamma$~\cite{persson85a,volokitin06a}. The oscillating 
probe excites phonons, whereby 
its motion is damped. The  solution of this problem proceeds via the Green's 
function of~\cref{eq:eqm} (SM).

\Cref{eq:eqm} is a widely applicable model of phonon dynamics 
\cite{findley13a,landau86a,lee55a}, and related models have been influential in 
understanding friction \cite{persson85a,hu20a}. It contains two fundamental 
modes with  longitudinal,
$\cl(\omega)=\sqrt{\frac{3K+4\mu-i\omega(3\xi+4\eta)}{3\rho}}$, and transverse, 
$
\ct(\omega)=\sqrt{\frac{\mu-i\omega\eta}{\rho}}$, speeds of sound.  
  $K$ and $\mu$ ($\xi$ and $\eta$) are the bulk and shear elastic (viscous) 
moduli, respectively, and $\rho$ the mass density. $\ct$ and $\cl$ are complex 
due to finite viscous moduli, i.e.,~\cref{eq:eqm} contains phonon attenuation. 
Microscopically, such attenuation may be caused by  phonon-phonon, 
phonon-electron, or phonon-defect scattering. For simplicity, we 
assume ${\cl(\omega)}/{\ct(\omega)}=\sqrt{3}$, independent of $\omega$, a good 
approximation for solids~\cite{persson85a,gornall71a,petert73a,cong19a}. $\cl$ 
thus drops out 
of the discussion.

The bulk situation, where the coating thickness $d$ in  Fig.~\ref{fig:drawing} 
is infinite, was 
extensively studied  previously~\cite{persson85a,persson99a,volokitin06a, 
lee21a}. In this case,
the probe excites spherical waves $\sim  e^{i \omega \rp/\ctz}/\rp$ (SM), and 
the damping coefficient reads
\begin{equation}
\begin{split}
    \Gamma^{\infty}&=\frac{\kappa^2\na^2l^4}{ \ctz^{\prime 3}\rho_0}\left(\zela 
    + \frac{\zvis\eta_0}{\ctz^\prime 
    \rho_0l}+\order{\frac{1}{Q^{2}},\frac{l^2}{\lambda^2}}\right),
\label{eq:FT_inf}
\end{split}
\end{equation}
with $\zela\approx1.29$,  $\zvis\approx1.72$, and $\ctz'$ the real part of 
$\ctz$. \cref{eq:FT_inf} is valid for $l\ll \lambda$, with phonon wavelength 
$\lambda=2\pi\ctz'/\omega$. For typical speeds of sound and $l$ in the 
nanometer range, the corrections to \cref{eq:FT_inf} are small for $\omega\alt 
\SI{e9}{\hertz}$. This number, $\SI{e9}{\hertz}$, is large compared to typical frequencies excited by the probe, since they are 
expected to be in the range 
$\omega\sim\SI{e2}{}\dots\SI{e5}{\hertz}$~\cite{dorofeyev99a,stipe01a_experiment,gotsmann01a_experiment,weber21a,schmidt20a,filleter09a,filleter10a,lee10a_gp}.
Additionally, \cref{eq:FT_inf} assumes the  quality factor  
$Q(\omega)=\frac{\mu_0}{\omega\eta_0(\omega)}$, the ratio between phonon decay 
length and phonon wave length, to be large compared to unity. This seems a 
justified assumption as well, as seen in estimates of the order of 
$Q\sim10^4$~\cite{ono20a,persson01a} (see Table~\ref{tab:param} for 
experimental parameters). Despite $Q$ being large, phonon attenuation is 
essential for friction since, as shown below, it dominates the behavior in 
certain regimes.

\begin{table}[b]
\begin{threeparttable}[b]
\begin{tabular*}{0.45\textwidth}{@{\extracolsep{\fill}}SSSS}\toprule
$l$~\tnote{a}&$\omega$~\tnote{b}&$\rho$~\tnote{c}&$\mu$~\tnote{c}\\ \midrule
\SI{1}{\nano\meter}&\SI{e3}{\per\second}&\SI{e-23}{\kilo\gram\per\nano\meter\cubed}&\SI{10}{\kilo\gram\per\nano\meter\per\second\squared}\\
 \bottomrule
\end{tabular*}
\begin{flushleft}
\footnotesize{$^a$ Refs.~\cite{weber21a,lee10a_gp,smolyanitsky12a}. $^b$ 
Refs.~\cite{weber21a,schmidt20a,filleter09a,filleter10a,lee10a_gp}. $^c$ 
Ref.~\cite{auld73a}.}
\end{flushleft}
\caption{Parameters typical for AFM experiments on solids:  radius of contact 
$l$, frequency $\omega$,  mass density $\rho$, and shear elastic modulus $\mu$.}
\label{tab:param}
\end{threeparttable}
\end{table}

The two leading terms in~\cref{eq:FT_inf} have fundamentally different physical 
origins. The first, called the elastic contribution, corresponds to transport 
of 
energy by traveling waves, and it persists  without phonon 
attenuation~\cite{persson85a,lee21a}.
The second, called the viscous contribution, is proportional to 
viscosity $\eta_0$. We discuss its physical origin below.
Their relative weight is the dimensionless viscosity,
\begin{equation}
\neta=\frac{\eta_0}{\ctz'\rho_0 l}=\frac{\lambda}{2\pi Q l},
\label{eq:num}
\end{equation}
which depends on  material properties and probe size $l$.
The numbers of~\cref{tab:param} for typical AFM conditions imply 
$\neta\sim\num{e5}$.  We thus  first 
discuss the case of $\neta\gg1$, where the bulk case of Eq.~\eqref{eq:FT_inf} 
is dominated by the viscous contribution (using $\ctz' = 
\sqrt{\mu_0/\rho_0}+\order{Q^{-2}}$)
\begin{equation}
\begin{split}
    \lim_{\neta \gg 
    1}\Gamma^{\infty}&=\frac{\zvis\kappa^2\na^2l^3\eta_0}{\mu_0^2}.
\end{split}
\label{eq:FT_vis}
\end{equation}
For finite values of $d$, we note that the boundary condition at the 
interface between coating (labeled with the subscript $0$) and substrate 
(subscript $1$) is determined by the refractive index 
$n(\omega)=\ctz(\omega)/\cto(\omega)$ and  $\rho_0/\rho_1$
\cite{brekhovskikh80a}. It is insightful to start with the limit 
$n=0$~\footnote{For $n=0$, $\rho_0/\rho_1$ is irrelevant.}, which yields a 
Dirichlet boundary condition (DBC) of  
$\bm{u}(\vrp,-d,\omega)=\bm{0}$~\cite{landau86a,brekhovskikh80a}.
In this case the substrate is much stiffer than the coating, and phonons are 
totally reflected at the interface, with phase shift $\pi$.

\begin{figure}
    \centering
    \includegraphics[width=0.4\textwidth]{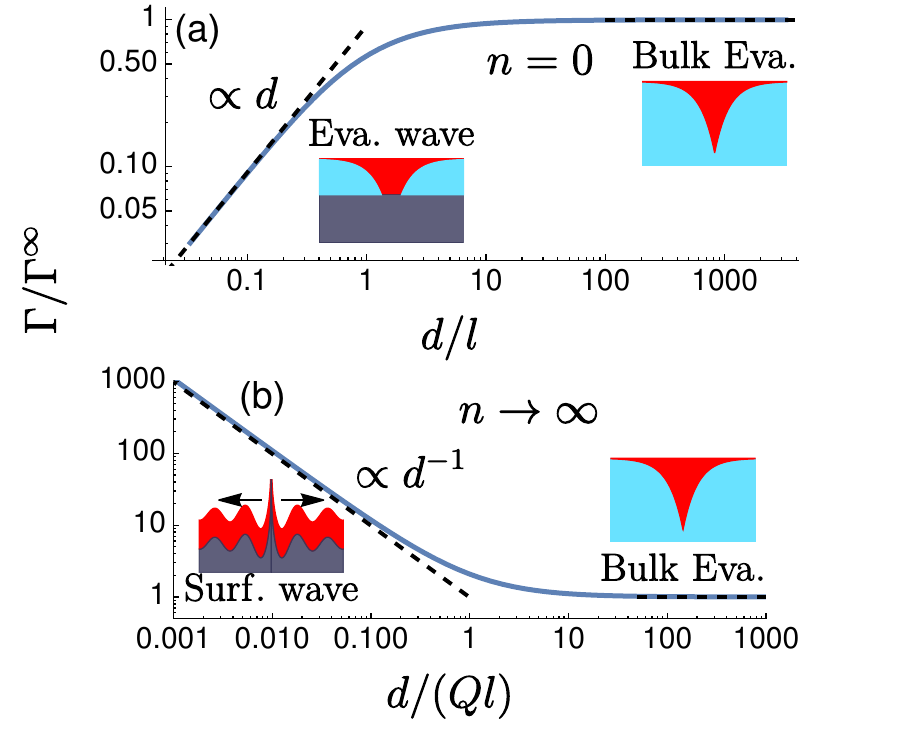}
    \caption{Damping coefficient $\Gamma$ for (a) $n=0$ and (b) $n\to\infty$ as 
    a function of (scaled) distance $d$, in the viscous limit $\neta\gg 1$. The 
    solid line shows numerical evaluation, dashed lines give the asymptotes of 
    Eqs.~\eqref{eq:FT_vis}, \eqref{eq:FT_lim1} and \eqref{eq:FT_lim2}, 
    respectively. Sketches give the fundamental wave solutions in the 
    corresponding regimes.
    }
    \label{fig:OF_si}
\end{figure}

\Cref{fig:OF_si} (a) shows $\Gamma$ as a function of $d$ for $n=0$ and 
$\neta\gg 1$, growing linearly in $d$,  and saturating to the 
bulk value of~\cref{eq:FT_vis} for large $d$, with a cross-over length scale 
around $d\approx l$.
Indeed,
we find that, for small $d$, the leading order of $\Gamma$ is linear in $d$,
\begin{equation}
    \Gamma=
         \frac{\pi\kappa^2\na^2l^2\eta_0 
         d}{2\mu_0^2}+\order{d^3}\stackrel{\tilde{\eta}\gg 1}{=} 
         \frac{\pi\Gamma^{\infty}}{2\zvis}\frac{d}{l}+\order{d^3}\hspace{5mm}\text{(DBC)}.
    \label{eq:FT_lim1}
\end{equation}
$\Gamma$ {\it vanishes} as $d\to0$, because the coating, when placed on a 
stiff substrate, supports fewer phonons as $d\to0$. In the second step 
in~\cref{eq:FT_lim1}, we used Eq.~\eqref{eq:FT_vis} to replace $\Gamma^\infty$, 
making apparent the mentioned saturation to $\Gamma^\infty$ at $d\approx l$. 
The dependence of $\Gamma$ on $d$ is {\it short range}, set by probe size $l$ 
of Eq.~\eqref{eq:vfps}.

How is this short range decay possible over distances much smaller than the 
wavelength? The motion of the probe excites not only (attenuated) traveling 
waves, but also \textit{evanescent waves} that decay within a range of $l$. In 
the presence of finite viscosity, they contribute to energy absorption, and 
thus to the damping coefficient $\Gamma$. In Fig.~\ref{fig:OF_si}, this 
mechanism outweighs the energy transported by traveling waves, which is why 
\cref{eq:FT_vis,eq:FT_lim1} carry $\eta_0$ as a factor.

The opposite limit of boundary conditions, $n\to\infty$, corresponds to a 
freestanding  coating or a much more compliant substrate,
and the waves obey a Neumann BC (NBC), 
$\bm{\hat{e}_z}\cdot\bm{\sigma}(\vrp,-d,\omega)=\bm{0}$ with stress tensor 
$\bm{\sigma}$ and surface normal $\bm{\hat{e}_z}$ \cite{brekhovskikh80a}. 
Phonons are totally reflected without 
phase shift.
This renders $\Gamma$ fundamentally different from the case 
of $n=0$ as shown in~\cref{fig:OF_si} (b); $\Gamma$ \textit{diverges} for 
small $d$ and converges to the bulk value on a scale of $d\approx Ql$. 
Expanding this case for small $d$ yields
\begin{equation}
    \Gamma=
     \frac{11\pi^2\kappa^2\na^2l^4}{64\mu_0\abs{\omega}d}+\order{d}\stackrel{\tilde{\eta}\gg
      1}{=}\frac{11\pi^2\Gamma^{\infty}}{64\zvis}  \frac{Ql}{d}+\order{d}
     \hspace{3mm}\text{(NBC)}.
\label{eq:FT_lim2}
\end{equation}
In this case, the probe excites \textit{traveling surface waves}, $\sim 
e^{i\omega \rp/\ctz}/\sqrt{\rp}$, i.e., the coating oscillates like a 
freestanding 2D sheet. Energy is transported along the surface, rather than 
absorbed; \cref{eq:FT_lim2} is independent of viscosity $\eta_0$.  As the  
sheet gets more compliant with $d\to 0$,  amplitudes of excitations get larger, 
and formally diverge as $d\to0$.
The second equality of Eq.~\eqref{eq:FT_lim2} implies  the mentioned saturation 
length of $d\approx   Ql$.
With $l$ on the scale of nanometers, this length is of the order of microns. It 
will be interesting to compare these surface modes to so-called 
puckering~\cite{lee10a_gp,li16a} or ploughing~\cite{smolyanitsky12b,deng12a} 
identified in previous work.

\begin{figure}
    \centering
    \includegraphics[width=0.48\textwidth]{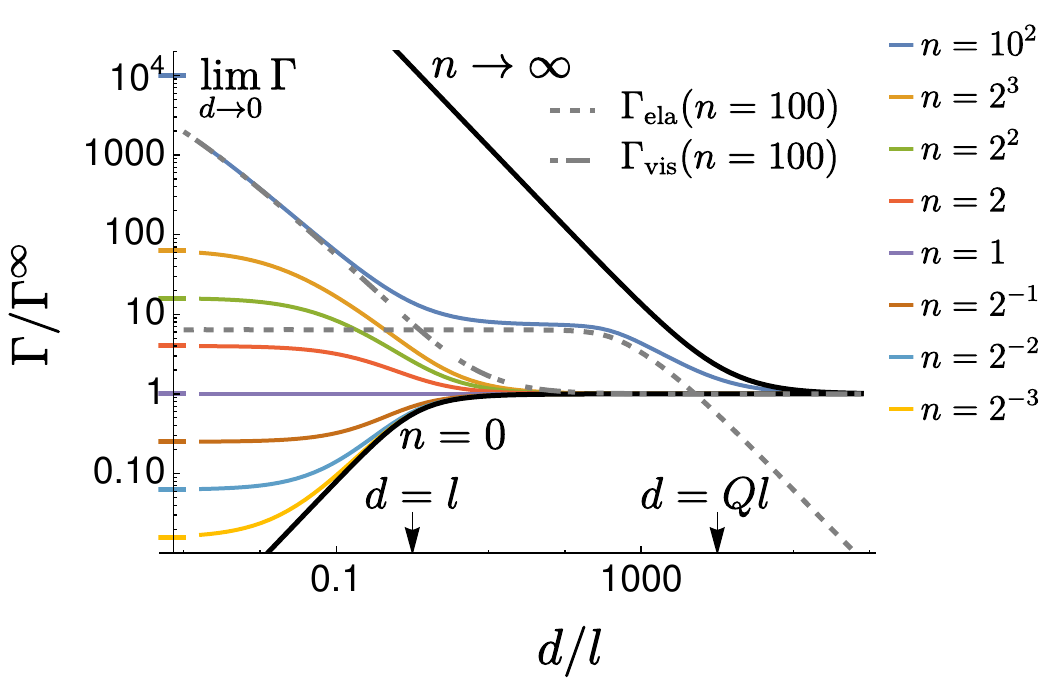}
    \caption{Damping coefficient $\Gamma$ for various refractive indices $n$, 
    using  $\neta=10^5$ and $Q=10^4$, only considering transverse waves. Marks 
    on the $y$ axis are (analytic) values of $\lim_{d\to0}\Gamma$ for the 
    corresponding $n$. The gray dashed lines are the purely elastic and viscous 
    contributions at $n=100$.}
    \label{fig:ref_index}
\end{figure}

The cases of $n=0$ and $n\to\infty$ provide a reference for the 
discussion of an arbitrary refractive index  $n$. For simplicity, 
we assume that the quality factors of the coating and substrate are identical, 
making $n(\omega)$ real, and use $\rho_0=\rho_1$ \footnote{These 
two assumptions can be relaxed without additional challenge, yielding similar 
conclusions as the ones presented.}.
For finite values of $n$, we note a numerical challenge in evaluating the 
longitudinal modes, so that we restrict the shown data to transverse waves (SM).
   
\Cref{fig:ref_index} shows $\Gamma$ as a function of $d$ for various $n$, at 
$\tilde\eta=10^5$ and $Q=10^4$; $\Gamma$ is monotonic in $d$, and stays within 
the bounds of the limiting cases of $n=0$ and $n\to\infty$. Importantly, for a 
thin coating, $d\to0$, $\Gamma$  approaches the bulk 
value of the substrate, indicated as bars on the $y$-axis. These are obtained 
by \cref{eq:FT_inf} with the material parameters of the 
substrate~\footnote{We assume that the surface coupling (here $\kappa$, $n_A$ 
and $l$) are the same for coating and for pure substrate.} \footnote{In 
experiments, the limit $d\to0$ (thin coating) is expected to be different from 
$d=0$ (no coating), because of different surface interactions for substrate and 
coating.}. 
$\Gamma$ thus varies between the bulk results of the substrate ($d\to0$) and of 
the coating ($d\to\infty$). 

The curves of Fig.~\ref{fig:ref_index} up to $n=2^3$ can be understood by these 
two limiting cases, and a transition on the length scale $l$. These cases are 
thus dominated by evanescent waves.
The dimensionless viscosity of the substrate equals $\tilde\eta/n$, and goes 
down for large values of $n$. The damping coefficient of the bulk substrate, 
i.e., the behavior at small $d$, is thus dominated by traveling waves at large 
$n$.
Indeed, for $n\to\infty$, the behavior of Fig.~\ref{fig:OF_si} b) is 
approached.  For intermediate values of $n$ ($n=10^2$ in the graph), a two-step 
decay occurs, with evanescent waves for $d\alt l$ and $d\agt Ql$,  and 
traveling surface waves for $l\alt d\alt Ql$.

\begin{figure}
    \centering
    \includegraphics[width=0.45\textwidth]{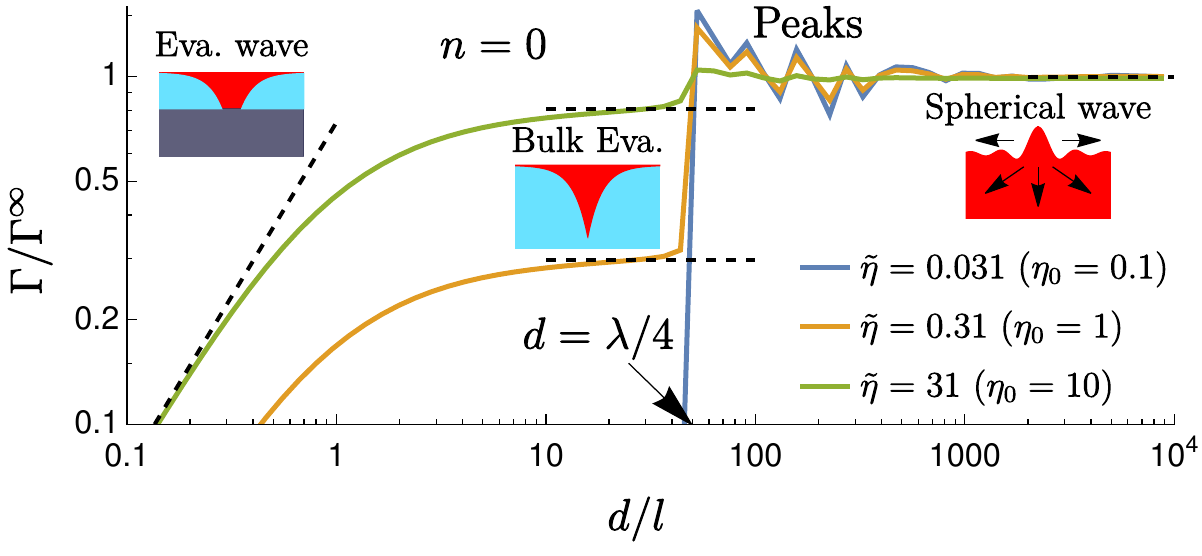}
\caption{Damping coefficient $\Gamma$ as a function of (scaled) distance for 
$n=0$, for three values of $\tilde\eta$. The black dashed lines represent the 
asymptotes of~\cref{eq:FT_lim1,eq:FT_vis,eq:FT_inf}.
For the chosen parameters, the first peak is at $d/l=\lambda/4l \approx 50$.
}
\label{fig:OF_au}
\end{figure}

What about the case of $\neta\ll1$  where the limit $d\to\infty$ is dominated 
by traveling waves?
Despite less experimental relevance to AFM experiments,
we include this insightful case in Fig.~\ref{fig:OF_au} for completeness, 
focusing on $n=0$ (see SM for $n\to\infty$).
As seen in the Figure,  
for $d\ll \lambda$ the curves are similar to Fig.~\ref{fig:OF_si} (a). This is 
because, for $n=0$, 
the coating {\it does not support} traveling waves for $d\alt\lambda$ (SM).
For $d\agt\lambda$, traveling waves are excited, yielding a sharp transition 
between evanescent and traveling waves at  $d\approx\lambda/4$, followed by 
peaks. These are due to 
interference effects between outgoing and reflected traveling waves (SM).
For $d\gg\lambda$, the bulk value of~\cref{eq:FT_inf} is approached.

\begin{figure}
    \centering
    \includegraphics[width=0.43\textwidth]{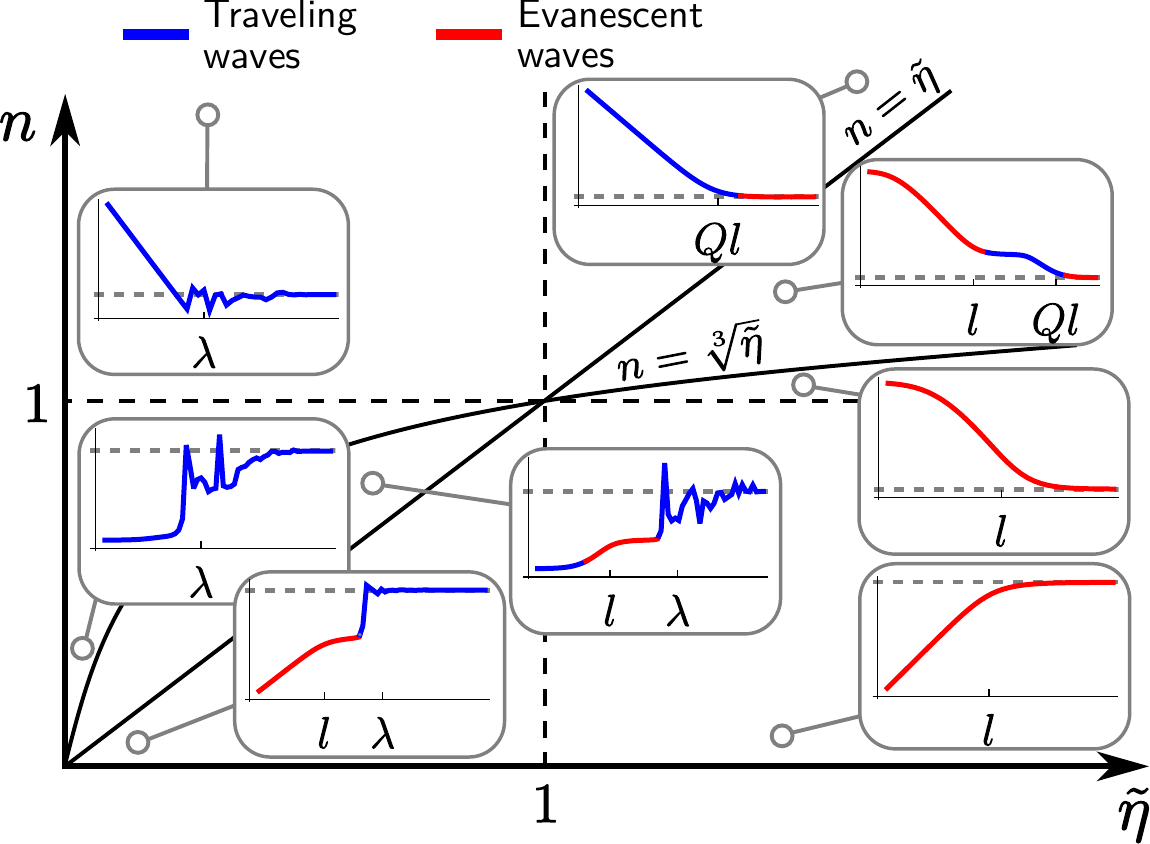}
    \caption{Friction map. Depending on $\tilde\eta$ and $n$, $\Gamma$ is 
    dominated by evanescent modes (red) and decays to bulk on a scale 
    $d\approx l$, or by traveling waves (blue), which decay on scales of $Ql$ 
    or $\lambda$. The lines separating the regimes are deduced from 
    Eq.~\eqref{eq:fric_sub}. Gray dashed lines show the limit $d\to\infty$ of 
    \cref{eq:FT_inf}.
    }
    \label{fig:phase}
\end{figure}

This discussion allows identification of mechanism regimes, depicted in 
Fig.~\ref{fig:phase}. The limit of $d\to\infty$ is dominated by evanescent 
waves for $\tilde\eta\gg1$, and by traveling waves for $\tilde\eta\ll1$. This 
is indicated in~\cref{fig:phase} by using different colors in the inset graphs. 
The line $n=\tilde\eta$ separates the same for $d\to0$, i.e., curves that begin 
with blue or red. This limit can be found  from   
\footnote{See the SM for more discussion and the general case where $n$ is 
complex and $\rho_0\ne\rho_1$.},
\begin{equation}
\lim_{d\to 0}\Gamma= \Gamma^\infty
n^3 \frac{\zela+\zvis\frac{\tilde \eta}{n}}{\zela+\zvis\tilde \eta}.
\label{eq:fric_sub}
\end{equation}
Another separator is $n=1$, dividing curves where $\Gamma^\infty$ is larger 
than the limit of $d\to 0$ from the opposite. The last curve, 
$n=\sqrt[3]{\tilde\eta}$, together with $n=\tilde\eta$, bounds two regimes with 
multiple transitions between traveling and evanescent waves. The latter occur 
because of the different $d$-dependence of traveling and evanescent waves.

This framework identifies the main phononic mechanisms for 
damping between a probe and a coated planar surface, with vastly different 
behaviors. These regimes are expected to 
occur in experiments and simulations, whenever the thickness of coatings can be 
changed without changing contact mechanics. Fig.~\ref{fig:phase} illustrates that pronounced dependence on layer thickness is expected, e.g., if the two materials show a refractive index very different from unity. The discussed isolated frequencies 
apply in non-contact measurements, where narrow frequency bands are excited 
\cite{dorofeyev99a,stipe01a_experiment,gotsmann01a_experiment,kisiel11a_experiment}.
 In sliding experiments, one expects a spectrum of frequencies, so that the 
results reported here need to be averaged accordingly. It is important to note 
that the contributions by evanescent waves are frequency independent within the 
range given below Eq.~\eqref{eq:FT_inf}, making this average trivial. Friction 
from traveling waves has frequency dependent features, see 
Eq.~\eqref{eq:FT_lim2}, manifested as the peaks in~\cref{fig:OF_au}.

The presented model is simple and provides analytical results, which we expect to improve understanding of friction phenomena. The quantitative translation to the case of sliding motion needs to be investigated in future work~\cite{weber22a}. We note however qualitative agreement with studies on 
graphene. Experiments~\cite{lee10a_gp} and simulations~\cite{smolyanitsky12a}  
reported friction on freely standing graphene flakes, i.e., $n\gg1 $, finding a 
decrease with  increasing number of layers. The same trend is observed for 
mounted graphene layers (or 
flakes)~\cite{filleter09a,filleter10a,lee10a_gp,smolyanitsky12a}; graphene is 
much stiffer ($\mu\sim\SI{1}{\tera\pascal}$~\cite{lee12a_gp,van00a,cong19a}) 
than metal substrates ($\mu\sim\SI{10}{\giga\pascal}$~\cite{auld73a,zhao99a}), 
so that the case of $n\gg1$ applies too. Previous work attributed this behavior 
to the enhanced rigidity for thicker samples~\cite{lee10a_gp,smolyanitsky12a}, 
leading to the suppression of surface waves, in agreement  with our findings.

The framework of Eq.~\eqref{eq:eqm} naturally includes Hertzian contact 
theory~\cite{landau86a,johnson82a_hertz} in the limit of small frequencies. It 
is interesting to remark that the identified evanescent waves can equally be 
found from the slowly moving distortion field of contact theory. Moving this 
distortion field dissipates energy due to viscosity (SM). This way, this approach  
 can be linked 
to a variety of other approaches based on contact 
theory~\cite{persson03a,hu20a,qu20a,weber22a} and local deformations 
\cite{smolyanitsky12b,li16a,smolyanitsky15a}. Natural extensions include 
nonlinear systems, where, e.g., deeper indentations 
\cite{smolyanitsky12b,li16a,smolyanitsky15a} can be studied. Lateral confinement was also found to be of importance \cite{wada18a}, and it provides an interesting additional possibility to identify involved phonon modes. This can be address in the presented framework in future work.

This work was funded by the Deutsche Forschungsgemeimschaft (DFG, German 
Research Foundation) - 217133147 via
SFB 1073 (Project A01). We thank Richard L.C. Vink for stimulating discussions. 
In memory of Philip Rauch.

\bibliography{bibtex/miru}
\bibliographystyle{apsrev4-2}
\end{document}


\preprint{APS/123-QED}
\title{Supplementary Material}
\author{Miru Lee}
\email{miru.lee@uni-goettingen.de}
\affiliation{Institute for Theoretical Physics, Georg-August-Universit\"at 
G\"ottingen, 37073 G\"ottingen, Germany}
\author{Niklas Weber}
\affiliation{Institute of Materials Physics, Georg-August-Universit\"at 
G\"ottingen, 37073 G\"ottingen, Germany}
\author{Cynthia A. Volkert}
\affiliation{Institute of Materials Physics, Georg-August-Universit\"at 
G\"ottingen, 37073 G\"ottingen, Germany}
\author{Matthias Kr\"uger}
\email{matthias.kruger@uni-goettingen.de}
\affiliation{Institute for Theoretical Physics, Georg-August-Universit\"at 
G\"ottingen, 37073 G\"ottingen, Germany}

\date{\today}
\maketitle

\section*{Damping coefficient}
The damping coefficient can be calculated from a Green-Kubo 
relation~\cite{kubo12a,zwanzig01a,kruger16a,lee21a},
\begin{equation}
\Gamma=\frac{1}{2 \kb 
T}\int_{-\infty}^{\infty}\frac{\dd{\omega'}}{2\pi}\avg{F(X,\omega) ; 
F(X,\omega')},
\end{equation}
where $\avg{A ; B}=\avg{(A-\avg{A})(B-\avg{B})}$ is an ensemble average of the 
covariance.
With the force defined in [Eq.~(1)] in the main text, the damping coefficient 
can be rewritten as
\begin{equation}
\begin{aligned}
\Gamma=&\frac{\kappa^2\na^2}{2\kb 
T}\intinf\dd[2]{\vrp}\intinf\dd[2]{\vrp'}\intinf\frac{\dd{\omega'}}{2\pi}e^{-\frac{\vrp^2}{l^2}}\avg{u_x(\vrp,\omega)u_x(\vrp',\omega')}e^{-\frac{\vrp'^2}{l^2}}\\
=& 
	\frac{\kappa^2n_A^2}{\omega} 
	\intinf\dd[2]{\bm{\rp}}\intinf\dd[2]{\bm{\rp}'}e^{-\frac{\vrp^2}{l^2}}\Im\{G(\vrp,\vrp',\omega)\}e^{-\frac{\vrp'^2}{l^2}}\\
=& 
	\frac{\kappa^2n_A^2}{\omega}\Im\bigg\{ 
	\intinf\dd[2]{\bm{\rp}}\intinf\dd[2]{\bm{\rp}'}e^{-\frac{\vrp^2}{l^2}}G(\vrp,\vrp',\omega)e^{-\frac{\vrp'^2}{l^2}}\bigg\}
\end{aligned}
\label{eq:fric}
\end{equation}
Note that in the second equality, we make use of the 
FDT~\cite{eckhardt84a,kruger11a,kruger12a,landau13a_stat_mech2,agarwal75a,lee21a},
\begin{equation}
\int_{-\infty}^{\infty}\frac{\dd{\omega'}}{2\pi}\avg{u_x(\bm{r},\omega)u_x(\bm{r}',\omega')}=\frac{2\kb
 T}{\omega}\Im{G(\bm{r},\bm{r}',\omega)}.
\end{equation}
It is more convenient to obtain the Green's function in $\vkp=(k_x,k_y)$ space, 
since the Kelvin-Voigt model ([Eq.~(2)] in the main text) becomes an ordinary 
differential equation.
Let us change the spatial integral to an integral over $\vkp$ space,
\begin{equation}
\begin{split}
\Gamma=& 
	\frac{\kappa^2n_A^2}{\omega}\Im\bigg\{ 
	\intinf\dd[2]{\bm{\rp}}\intinf\dd[2]{\bm{\rp}'}e^{-\frac{\vrp^2}{l^2}}G(\vrp,\vrp',\omega)e^{-\frac{\vrp'^2}{l^2}}\bigg\}\\
	=& 
	\frac{\kappa^2n_A^2}{\omega}\Im\bigg\{ \int\dd[2]{\vrp}\int\dd[2]{\vrp'}
	\intinf\frac{\dd[2]{\bm{\kp}}}{(2\pi)^2}
	\intinf\frac{\dd[2]{\bm{\kp}'}}{(2\pi)^2} 
	e^{-\frac{\vrp^2}{l^2}}G(\vkp,z=0,\vkp',z'=0,\omega)e^{-\frac{\vrp'^2}{l^2}}
	e^{i\vkp\cdot\vrp}e^{i\vkp'\cdot\vrp'}\bigg\}\\
	=&\frac{\kappa^2n_A^2 l^4\pi^2}{\omega}\Im\bigg\{
	\intinf\frac{\dd[2]{\vkp}}{(2\pi)^2}
	\intinf\frac{\dd[2]{\vkp'}}{(2\pi)^2}
	e^{-\frac{\kp^2 l^2}{4}}G(\bm{\kp},0,\bm{\kp}',0,\omega)
	e^{-\frac{\kp'^2 l^2}{4}}\bigg\}.
\end{split}
\label{eq:fric_k_omega0}
\end{equation}
The above expression for the damping coefficient can be further simplified if 
the homogeneity of the Green's function is assumed, i.e., 
$G(\bm{r}+\bm{r}_0,\bm{r}'+\bm{r}_0,\omega)=G(\bm{r},\bm{r}',\omega)$~\cite{vliet08a},
\begin{equation}
\begin{split}
G(\bm{k},\bm{k}',\omega)&=\intinf\dd[3]{\bm{r}}\intinf\dd[3]{\bm{r}'}G(\bm{r}+\bm{r}_0,\bm{r}'+\bm{r}_0,\omega)e^{-i\bm{k}\cdot(\bm{r}+\bm{r}_0)}e^{-i\bm{k}'\cdot(\bm{r}'+\bm{r}_0)}\\
&=\intinf\dd[3]{\bm{r}}\intinf\dd[3]{\bm{r}'}G(\bm{r},\bm{r}',\omega)e^{-i\bm{k}\cdot\bm{r}}e^{-i\bm{k}'\cdot\bm{r}'}e^{-i\bm{r}_0\cdot(\bm{k}+\bm{k}')}\\
&=G(\bm{k},\bm{k}',\omega)e^{-i\bm{r}_0\cdot(\bm{k}+\bm{k}')}.
\end{split}
\end{equation}
The homogeneity thus entails $\bm{k}=-\bm{k}'$, i.e.,
\begin{equation}
G(\bm{k},\bm{k}',\omega)=(2\pi)^3G(\bm{k},\omega)\delta(\bm{k}+\bm{k}').
\end{equation}
Or,
\begin{equation}
	G(\bm{\kp},z,\bm{\kp}',z',\omega)=(2\pi)^2G(\bm{\kp},z-z',\omega)\delta(\bm{\kp}+\bm{\kp}').
\end{equation}
Plugging it into~\cref{eq:fric_k_omega0}, one arrives at
\begin{equation}
\begin{split}
\Gamma =&\frac{\kappa^2n_A^2 l^4 \pi^2}{\omega}\Im\bigg\{
\intinf\frac{\dd[2]{\bm{\kp}}}{(2\pi)^2}G(\bm{\kp},0,\omega)e^{-\frac{1}{2}\kp^2l^2}\bigg\},
\end{split}
\label{eq:fric_k_omega}
\end{equation}
which we use to evaluate the damping coefficient both numerically and 
analytically.

\section*{Boundary conditions}
Regarding the geometry of the system, it has the surface at $z=0$ in the $xy$ 
plane, while extending infinitely in the $x$ and $y$ directions. For $z>0$, it 
is a vacuum, i.e., no phonon can exist. The boundary conditions (BC) at $z=0$ 
are given by~\cite{landau86a,findley13a,lee55a,mindlin36a}
\begin{equation}
\begin{aligned}
\sigma_{xz}(\vrp,\omega)=\kappa X(\omega)e^{-\frac{\vrp^2}{l^2}},\\
\end{aligned}
\label{eq:st_bound}
\end{equation}
while other components are zero.
Here, $\bm{\sigma}(\bm{r},\omega)$ is the Cauchy stress tensor, which is 
defined as
\begin{equation}
\bm{\sigma}=\rho(\cl^2-2\ct^2)\nabla\cdot\bm{u}+\rho\ct^2(\nabla\bm{u}+(\nabla\bm{u})^T).
\end{equation}
Note that the BC on the surface suggest that the displacement field 
$u_x(\bm{r},\omega)$ is linearly proportional to the position of the probe 
$X(\omega)$, containing the phase information with respect to $X(\omega)$, 
which, in turn, yields a finite damping coefficient.

For $-d<z\le0$, we have the first layer (the coating) of the solid whose 
properties are characterized by $\clz(\omega)$, $\ctz(\omega)$, and $\rho_0$.
The second layer (the substrate) expands for $-\infty<z<-d$ with 
$\clo(\omega)$, $\cto(\omega)$, and $\rho_1$.
The BC at $z=-d$ require that the displacement field and traction are 
continuous~\cite{brekhovskikh80a},
\begin{equation}
\begin{aligned}
\lim_{z\to-d^+}\bm{u}(\bm{r},\omega)&=\lim_{z\to -d^-}\bm{u}(\bm{r},\omega),\\
\lim_{z\to-d^+}\bm{\hat{e}_z}\cdot\bm{\sigma}(\bm{r},\omega)&=\lim_{z\to 
-d^-}\bm{\hat{e}_z}\cdot\bm{\sigma}(\bm{r},\omega),
\end{aligned}
\end{equation}
where $\bm{\hat{e}_z}$ is the unit vector in the $z$ direction.
The Dirichlet and Neumann BC in the main text are the special cases of the 
above BC.

\section*{Traveling vs. evanescent wave}
Here, we show that traveling waves yield the elastic contribution, and 
evanescent waves the viscous contribution. Since any wave can be expressed by a 
superposition of plane waves, an ansatz of the transverse motion of the solid 
([Eq.~(2)] in the main text) can be written as
\begin{equation}
G_{\mathrm{T}}(\vkp,z-z',d,\omega)=C_0(\vkp,d,\omega)e^{\qt 
(z-z')}+C_1(\vkp,d,\omega)e^{-\qt (z-z')},
\end{equation}
where $\qt=(\kp^2-\omega^2/\ct^2)^{1/2}$, and $C_{0}(\vkp,d,\omega)$ 
($C_{1}(\vkp,d,\omega)$) is an amplitude of outgoing (incoming) wave.
The longitudinal counterpart can be similarly written with 
$\ql=(\kp^2-\omega^2/\cl^2)^{1/2}$, which then gives us the full solution 
$G=G_{\mathrm{T}}+G_{\mathrm{L}}$. From the ansatz, it follows that if 
$\kp\le\omega/\ct$, the solution represents a traveling wave in the $z$ 
direction. Contrarily, if $\kp\ge\omega/\ct$, then it decays exponentially in 
the $z$ coordinate, i.e., an evanescent wave, resulting from an adiabatic 
deformation (see below).

Notice also that the integral in~\cref{eq:fric_k_omega} runs in $\vkp$ space, 
along which there may exist branches or singularities. Consequently, the 
integral can be divided into many parts according to the branch or/and singular 
points,
\begin{equation}
\begin{aligned}
\Gamma=&\frac{\kappa^2\na^2 
l^4}{4\omega}\Im\bigg\{\int_0^{2\pi}\dd{\theta}\bigg[\int_{0}^{k_{s,1}}\dd{\kp}\kp
 G(\vkp,0,d,\omega)e^{-\frac{1}{2}\kp^2l^2}+\int_{k_{s,1}}^{k_{s,2}}\dd{\kp}\kp 
G(\vkp,0,d,\omega)e^{-\frac{1}{2}\kp^2l^2}\\
&+\int_{k_{s,2}}^{k_{s,3}}\dd{\kp}\kp 
G(\vkp,0,d,\omega)e^{-\frac{1}{2}\kp^2l^2}+\dots+\int_{k_{s,N}}^{\infty}\dd{\kp}\kp
 G(\vkp,0,d,\omega)e^{-\frac{1}{2}\kp^2l^2}\bigg]\bigg\}
\end{aligned}
\end{equation}
where $k_{s,n}~n\in\{1,2,\dots,N\}$ are the branch and singular points of the 
Green's function. In what follows, we show that the terms running up to 
$k_{s,N}$ integrate traveling waves resulting in the elastic contribution, and 
the last term evanescent waves resulting in the viscous contribution.

Let us exemplarily consider the transverse mode.
The Green's function at $z=z'$ (the argument is henceforth omitted) at the 
limit of $d\gg\lambda$ reads,
\begin{equation}
G_{\mathrm{T}}(\bm{\kp},\omega)=\frac{\sin^2\theta}{\rho_0\ctz^2\qtz}.
\label{eq:GFT_inf}
\end{equation}
The branch point is located at $k_s=\omega/\ctz$. Note that this branch point 
coincides with the singularity.
The integration is thus divided into two parts. The first part integrates 
traveling waves,
\begin{equation}
\frac{\kappa^2\na^2 
l^4}{4\omega}\int_0^{2\pi}\dd{\theta}\int_{0}^{\omega/\ctz}\dd{\kp}\kp 
\frac{\sin^2\theta}{\rho_0\ctz^2\qtz}e^{-\frac{1}{2}\kp^2l^2}=-\frac{\pi^{3/2}\kappa^2\na^2l^3
 e^{-l^2\omega^2/(2\ctz^2)}}{2^{5/2}\ctz^2\rho_0\omega}\erf\left(\frac{-i 
l\omega}{\sqrt{2}\ctz}\right),
\end{equation}
where $\erf(x)$ is the error function.
Expanding it at a small $\omega$ and taking the imaginary part, one arrives at
\begin{equation}
\Im\left\{-\frac{\pi^{3/2}\kappa^2\na^2l^3 
e^{-l^2\omega^2/(2\ctz^2)}}{2^{5/2}\ctz^2\rho_0\omega}\erf\left(\frac{-i 
l\omega}{\sqrt{2}\ctz}\right)\right\}=\frac{\pi 
\kappa^2\na^2l^4}{4}\sqrt{\frac{\rho_0}{\mu_0^3}}+\order{\frac{1}{Q^2},\frac{l^2}{\lambda^2}}.
\end{equation}
Note that one should perform the integral first and then the expansion, 
since the singularity renders these operations non-commute.
Note also that the leading order of this contribution is \textit{independent} 
of viscosity $\eta_0$.

The second part, on the contrary, takes care of the evanescent wave solution
\begin{equation}
\begin{aligned}
&\frac{\kappa^2\na^2 
l^4}{4\omega}\int_0^{2\pi}\dd{\theta}\int_{\omega/\ctz}^{\infty}\dd{\kp}\kp 
\frac{\sin^2\theta}{\rho_0\ctz^2\qtz}e^{-\frac{1}{2}\kp^2l^2} \\ &\qquad= 
\frac{\pi\kappa^2\na^2 
l^4}{4\omega}\int_0^{2\pi}\dd{\theta}\int_{0}^{\infty}\dd{\kp}\kp\sin^2\theta
\left[\frac{1}{\kp\mu_0}+\frac{i\omega\eta_0}{\kp\mu_0^2}\right]e^{-\frac{1}{2}\kp^2l^2}+\order{\frac{1}{Q^2}}.
\end{aligned}
\label{eq:eva_integral}
\end{equation}
Here, we expanded the integrand for small $\omega$ first, since the integral 
over $\kp$ and the limit of $\omega\to0$ commute~\cite{lee21a}.
Performing the integration leads us to
\begin{equation}
\frac{\pi\kappa^2\na^2 
l^4}{4\omega}\int_0^{2\pi}\dd{\theta}\int_{0}^{\infty}\dd{\kp}\kp\sin^2\theta
\left[\frac{1}{\kp\mu_0}+\frac{i\omega\eta_0}{\kp\mu_0^2}\right]e^{-\frac{1}{2}\kp^2l^2}
= \frac{\pi^{3/2}\kappa^2\na^2 l^3}{2^{5/2}\omega}
\left[\frac{1}{\mu_0}+\frac{i\omega\eta_0}{\mu_0^2}\right].
\end{equation}
Taking the imaginary part, one finds the friction contribution from evansecent 
waves
\begin{equation}
\frac{\pi^{3/2}\kappa^2\na^2l^3\eta_0}{2^{5/2}\mu_0^2}+\order{\frac{1}{Q^2}}.
\end{equation}
Unlike the elastic contribution from traveling waves, the leading order is 
linearly proportional to the viscosity $\eta_0$.
Putting them together, we find
\begin{equation}
\begin{split}
\Gamma_{\mathrm{T}}^{\infty}&= \frac{\pi 
\kappa^2\na^2l^4}{4}\sqrt{\frac{\rho_0}{\mu_0^3}} + 
\frac{\pi^{3/2}\kappa^2\na^2l^3\eta_0}{2^{5/2}\mu_0^2} 
+\order{\frac{1}{Q^2},\frac{l^2}{\lambda^2}}\\
&=\frac{\pi\kappa^2\na^2l^4}{4\rho\ctz^{\prime 3}}\left(
1+\sqrt{\frac{\pi}{2}}\frac{\eta_0}{\rho_0\ctz' l}\right)
+\order{\frac{1}{Q^2},\frac{l^2}{\lambda^2}}.
\end{split}
\label{eq:fric_trans}
\end{equation}
It is noteworthy that the existence of a branch point or a singularity at a 
finite value of $\kp$ directly indicates the existence of a traveling wave. If 
the Green's function exhibits no such point, then the corresponding wave is 
purely evanescent.

\section*{Evanescent waves and local deformation}
In this section, we show that evanescent waves represent the adiabatic local 
deformation due to the probe-sample interaction $F$, i.e., the Hertzian 
contact. The adiabatic local deformation can be obtained when the inertial term 
in [Eq. (2)] in the main text vanishes~\cite{landau86a,lee55a,mindlin36a}.
Exemplarily, let us consider the semi-infinite solid. The Green's function for 
the transverse mode reads
\begin{equation}
G_\mathrm{T}(\kp) = 
\sin^2\theta\left[\frac{1}{\kp\mu_0}+\frac{i\omega\eta_0}{\kp\mu_0^2}\right].
\end{equation}
This expression is precisely the Green's function in the rhs 
of~\cref{eq:eva_integral}. Note that since the probe-sample interaction acts 
parallel to the surface, this deformation represents the local shearing (e.g., 
puckering). Importantly, the local deformation contributes to the friction if 
and only if the solid is \textit{viscoelastic}. This is because it is the 
imaginary part of the Green's function that gives rise to the damping 
coefficient, which is non-zero only for viscoelastic solids.

\section*{Fundamental solutions}
For the Dirichlet BC, the Green's function at the surface ($z=z'=0$), for $d$ 
being the smallest length scale, is given by
\begin{equation}
G(\bm{\kp},d,\omega)=\frac{d}{\ctz^2\rho_0}+\order{d^3}.
\label{eq:G_fixed}
\end{equation}
The leading order has no branch or singular point, meaning that it is 
evanescent.
Since it is constant in $\vkp$ space, the above expression becomes a delta 
function in $\vrp$ space,
\begin{equation}
G(\vrp-\vrp',d,\omega)=\frac{d}{\ctz^2\rho_0}\delta^{(2)}(\vrp-\vrp')+\order{d^3}.
\end{equation}
For the Neumann BC, the Green's function reads
\begin{equation}
G(\bm{\kp},d,\omega)=\frac{\clz^2\left(\kp^2-\qlz^2\right)\left(k_y^2+\qtz^2\right)-2\ctz^2
 k_y^2\left(\kp^2-2\qlz^2+\qtz^2\right)}
{d\ctz^2\rho_0\qtz^2\left(\clz^2\left(
\kp^2-\qlz^2\right)\left(\kp^2+\qtz^2\right)-2\ctz^2\kp^2
\left(\kp^2-2\qlz^2+\qtz^2\right)\right)}+\order{d}.
\label{eq:G_open}
\end{equation}
One can easily tell it can be either a traveling or evanescent wave due to the 
existence of the singular point.
By finding the real space counterpart, one can easily identify the dimensional 
nature of the solution. Assuming $\ctz=\sqrt{3}\clz$, it is
\begin{equation}
G(\vrp-\vrp',d,\omega)=\frac{8(y-y')^2 
K_0\left(-\frac{i|\vrp-\vrp'|\omega}{\ctz}\right)+3(x-x')^2 
K_0\left(-\frac{i\sqrt{3}|\vrp-\vrp'|\omega}{2\sqrt{2}\ctz}\right)}{16\pi\rho_0\ctz^2
|\vrp-\vrp'|^2 d}+\order{d},
\end{equation}
where $K_0$ is the Bessel function of the second kind of the zeroth order, a 
form of  surface (2D) waves.

At a large $d\gg\lambda$ on the other hand, the Green's function is given by
\begin{equation}
G(\bm{\kp},\omega)=\frac{\clz^2\left(\kp^2-\qlz^2\right)\left(k_y^2+\qtz^2\right)-2\ctz^2
 k_y^2\left(\kp^2-2\qlz\qtz+\qtz^2\right)}
{\ctz^2\rho_0\qtz\left(\clz^2\left(
\kp^2-\qlz^2\right)\left(\kp^2+\qtz^2\right)-2\ctz^2\kp^2
\left(\kp^2-2\qlz\qtz+\qtz^2\right)\right)}.
\label{eq:Gxxinf}
\end{equation}
This, again, can be either a traveling or evanescent wave, however, finding the 
exact expression of the real space counterpart seems difficult.
A simpler way to see the dimensionality of the fundamental solution is to only 
consider the transverse mode, i.e.,~\cref{eq:GFT_inf}. In real space, it is 
given by
\begin{equation}
G_{\mathrm{T}}(\vrp-\vrp',\omega)=-\frac{(y-y')^2e^{\frac{i|\vrp-\vrp'|\omega}{\ctz}}}{2\pi\rho_0\ctz^2
 |\vrp-\vrp'|^3},
\end{equation}
which is a spherical (3D) wave solution.

Upon arriving at these fundamental wave solutions, no assumption on $\neta$ is 
made. In fact, $\neta$ can only be defined after obtaining~\cref{eq:Gxxinf}. 
This means, mathematically, that [Eqs. (6) and (7)] in the main text are valid 
for all $\neta$.

\section*{The damping coefficient for an arbitrary interaction at the limiting 
cases}
Let us consider a general way to formulate the probe-sample interaction $F$. 
Such an interaction arises from the underlying pairwise probe-atom interaction 
$g(\bm{X},\vrp,\omega)$ (it can easily be generalized for vectors as well),
\begin{equation}
F(X,\omega)=\na\intinf\dd[2]{\vrp}g(X,\vrp,\omega).
\label{eq:force}
\end{equation}
Expanding the pairwise interaction in $u_x$, one finds
\begin{equation}
g(X,\vrp,\omega)=g(X,\vrp)+ \pdv{x}g(X,\vrp) \cdot u_x(\vrp,\omega)+\cdots.
\end{equation}
If $\sqrt{\avg{u_x^2(\vrp,\omega)}}\ll\abs{\vrp-X}$ is assumed, the first term 
can be seen as the pairwise interaction in phase with the motion of $X$, and 
the second term out of phase. Now one can calculate the damping coefficient 
similarly as~\cref{eq:fric}.

Let us consider the case of small $d$.
For the Dirichlet BC, we use~\cref{eq:G_fixed} for the Green's function and 
perform the integral. Because the Green's function is independent of $\vkp$ for 
the leading order of $d$, the damping coefficient is given simply by the 
integration of the interaction,
\begin{equation}
\begin{aligned}
\Gamma =& \frac{\na^2 \eta_0 
d}{\mu_0^2}\intinf\frac{\dd[2]{\vkp}}{(2\pi)^2}\abs{\pdv{x}g(X,\vkp)}^2.
\end{aligned}
\end{equation}
The linear dependence of the damping coefficient on the coating thickness $d$ 
is thus universal to any type of interaction.

For the Neumann BC, the Green's function is given by~\cref{eq:G_open}, which, 
for $\ctz=\sqrt{3}\clz$, reduces to
\begin{equation}
G'(\kp,\omega)=\int_{0}^{2\pi}\dd{\theta}G(\vkp,\omega)
=\frac{\pi(11\ctz^2 \kp^2 - 
6\omega^2)}{d\rho_0(\ctz^2\kp^2-\omega^2)(8\ctz^2\kp^2-3\omega^2)}.
\end{equation}
Note that $g(X,\vkp)$ is azimuthally symmetric, so that it can be factored out 
of the integral over $\theta$.
The integral over $\kp$ can be done by means of the residue theorem due to the 
singularities at
\begin{equation}
\kp=\frac{\omega}{\ctz(\omega)},\qquad\kp=\sqrt{\frac{3}{8}}\frac{\omega}{\ctz(\omega)},
\end{equation}
yielding the elastic contribution from traveling waves. The remaining integrals 
that are associated with the residue theorem contributes to higher order terms 
in $\omega$ (i.e., $\order{d/\lambda}$, see Ref.~\cite{lee21a} for the detailed 
calculation).

The resulting damping coefficient for the leading order of $\omega$ and $d$ is
\begin{equation}
\Gamma=\frac{11}{64\mu_0\abs{\omega} d}\avg{F(X)}^2.
\end{equation}
The above expression is universal to any types of interaction for the leading 
order of $\omega$. In fact, the elastic contribution is agnostic to the form of 
interaction, the characteristic of which we named \textit{universality} in 
Ref.~\cite{lee21a}. To see this, let us consider the locations of the 
singularities at $\omega\to 0$; the singularities are approaching to $\kp=0$ as 
$\omega\to0$. This means the Green's function behaves like a delta function, 
and therefore the interaction has to be evaluated at $\kp=0$. 
By~\cref{eq:force}, $g(X,\vkp=\bm{0})=F(X)$ can be found; the damping 
coefficient does not depend on the form of the interaction. For detailed 
discussions and derivations of the universality of the elastic contribution, we 
refer the readers to Ref.~\cite{lee21a}.

\section*{Transverse vs. Full solutions}
\begin{figure}
    \centering
    \includegraphics[width=0.5\textwidth]{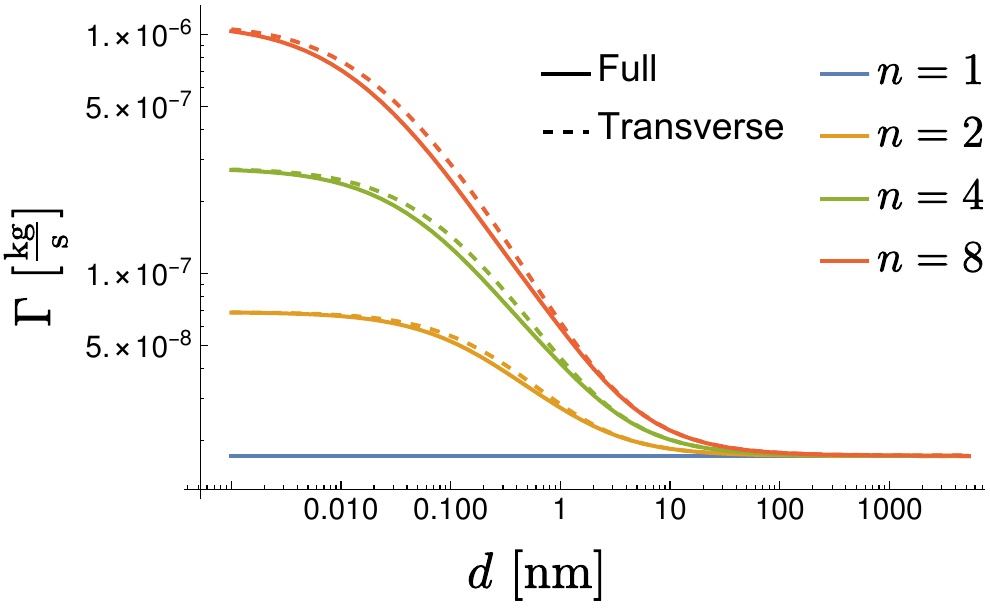}
    \caption{Comparison between the full (transverse and longitudinal) and 
    transverse solution calculations. $\Gamma$'s obtained from the transverse 
    solution are scaled by the factor in~\cref{eq:factor}.}
    \label{fig:index_comp}
\end{figure}
In the main text, the cases for arbitrary $n$ are calculated with only the 
transverse waves considered. In~\cref{fig:index_comp}, we present, for a few 
selected $n$ values, a comparison between the full (transverse and 
longitudinal) and the transverse wave solutions. Note that in the plot, the 
damping coefficients calculated from the transverse solution are scaled by a 
factor
\begin{equation}
\frac{\Gamma^\infty}{\Gamma_\mathrm{T}^\infty} = 
\frac{6.88\sqrt{2}}{\pi\sqrt{\pi}},
\label{eq:factor}
\end{equation}
which is nothing but the ratio of [Eq.~(5)] in the main text to the viscous 
contribution in~\cref{eq:fric_trans}. From~\cref{fig:index_comp}, it is clear 
that the transverse waves can capture the essential physics when it comes to 
the friction calculation.

\section*{Finding the peaks}
For $\neta \ll 1$, the damping coefficient $\Gamma$ exhibits peaks at 
$d\approx\lambda$, resulting from resonances of the excited waves. We identify 
that there are two types of resonances: body waves and surface waves. The 
former is due to interference of outgoing and reflected waves. The surface wave 
resonance, on the other hand, is related to the singularity of the Green's 
function in Fourier space 
($\vkp,\omega$)~\cite{persson85a,volokitin06a,persson99a}.

\begin{figure}
\subfloat{
    \centering
    \includegraphics[width=0.98\textwidth]{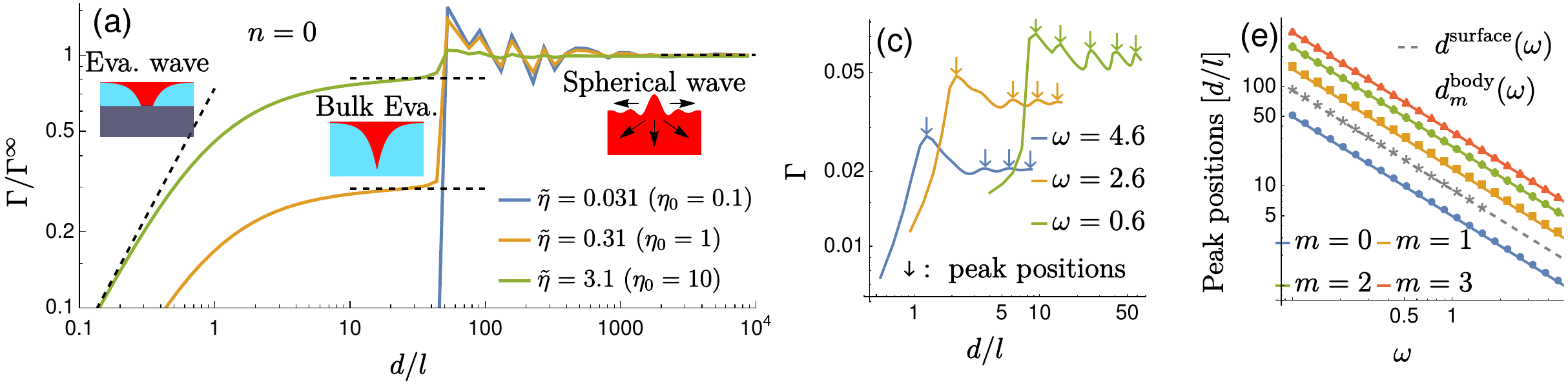}}\\
\subfloat{
    \centering
    \includegraphics[width=0.98\textwidth]{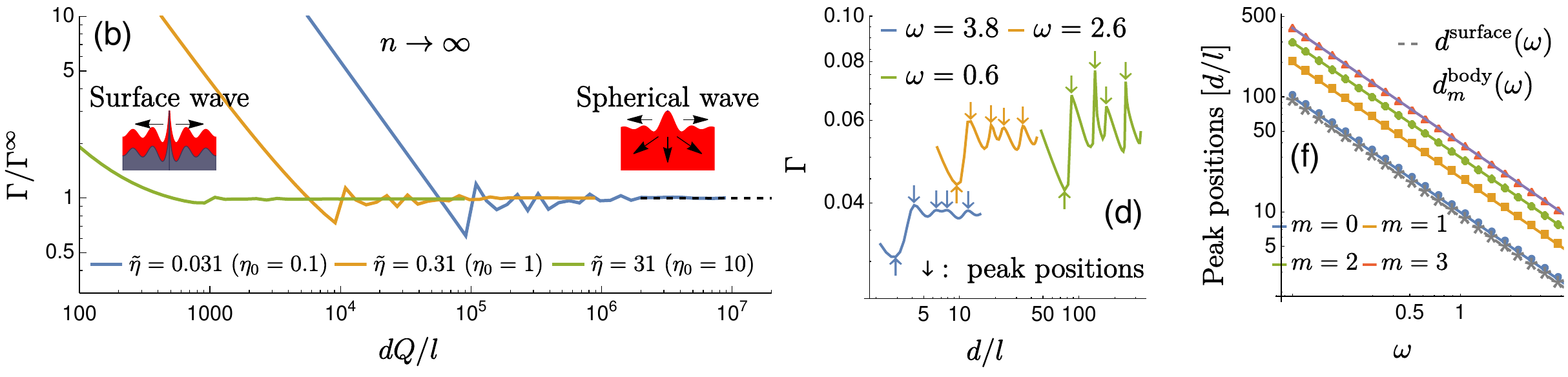}}
\caption{(a) and (b): The damping coefficient $\Gamma$ as a function of the 
depth of the first layer $d$ for $n=0$ (a) and $n\to\infty$ (b).
The solid lines are numerical evaluations of the friction, whereas the black 
dashed lines are the asymptotes. The parameters are 
$\kappa=1,~l=1,~\omega=0.1,~\na=1,~\rho=1,$ and $~\mu=10$, resulting in 
$\lambda/l\approx 200$.
(c) and (d): A closer look of the damping coefficient near the peak positions 
with different values of $\omega$ at $\neta=0.31$. (e) and (f): The first few 
peak positions as a function of $\omega$ at $\neta=0.31$. The points are the 
detected peak positions from the numerical evaluations, and the lines are the 
exact peak positions.}
\label{fig:open_au}
\end{figure}

Let us begin with the peak of the surface wave. From~\cref{eq:Gxxinf}, one 
finds that the position of singularity is at
\begin{equation}
    k_s(\omega)=\frac{\sqrt{3+\sqrt{3}}}{2}\frac{\omega}{\ctz(\omega)}.
\end{equation}
The peak position of surface wave is thus
\begin{equation}
    d^\mathrm{surface}(\omega)=\frac{1}{2}\Re\left\{\frac{2\pi}{k_s(\omega)}\right\}.
\end{equation}
The peaks of body waves can be found by considering the plane waves, i.e., 
evaluating the Green's function at $\vkp=\bm{0}$,
\begin{equation}
G(\bm{0},d,\omega)=
\begin{cases}
\frac{1}{\rho_0\ctz}\tan\left(\frac{d\omega}{\ctz}\right)&n=0\\
-\frac{1}{\rho_0\ctz}\cot\left(\frac{d\omega}{\ctz}\right)&n\to\infty,
\end{cases}
\end{equation}
from which the peak positions are obtained,
\begin{equation}
d^\mathrm{body}_m(\omega)=\begin{cases}
\frac{\pi\ctz'}{\omega}\left(\frac{1}{2}+m\right)&n=0\\
\frac{\pi\ctz'}{\omega}(1+m)&n\to\infty,
\end{cases}
~m\in\{0,1,2,\cdots\}
\end{equation}
Figures \ref{fig:open_au} (a) and (b) show the peak positions for $n=0$ and 
$n\to\infty$, respectively. The detected peak positions agree with our 
predictions as shown in the subsequent plots.
Note that in~\cref{fig:open_au} (b) we do not observe the plateau from the 
viscous contribution before the peaks. This is because $\neta$ and $Q$ are 
inversely related to each other, resulting in $l Q \approx \lambda$; the peaks 
occur before the first plateau of the viscous contribution fully establishes.

\section*{Defining the regimes in [Fig.~5] in the main text}
The scaling behavior of $\Gamma$ can be understood by studying 
$\lim_{d\to0}\Gamma$ and $\Gamma^{\infty}$,
\begin{equation}
\frac{\lim_{d\to0}\Gamma}{\Gamma^\infty} =  
\frac{n^3\zela+n^2\zvis\neta}{\zela+\zvis\neta}.
\label{eq:fric_sub}
\end{equation}
Keep in mind that $\zela/\zvis \approx 1$.
The numerator is the damping coefficient of the substrate and the denominator 
that of the coating. To begin with, let us consider the numerator, from which a 
separating line
\begin{equation}
n=\neta
\label{eq:line1}
\end{equation}
can be defined. If $n\gg \neta$, the elastic contribution of the substrate is 
much larger than the viscous contribution of the substrate.

Another line can be found when comparing the elastic contribution of the 
substrate with the viscous contribution of the coating,
\begin{equation}
n=\sqrt[3]{\neta}.
\label{eq:line2}
\end{equation}
If $\sqrt[3]{\neta}\alt n \alt \neta$, one finds $\zvis\neta \alt \zela n^3 $; 
the viscous contribution of the coating is smaller than the elastic 
contribution of the substrate. Because the latter decays at a slow rate 
$d\approx Q l$, it is visible as shown in [Fig.~3] for $n=100$ in the main 
text. Another case is when $\sqrt[3]{\neta} \alt n\alt 1$. The elastic 
contribution of the substrate is larger than the viscous contribution of the 
coating. In this case, the damping coefficient remains constant until $d\approx 
\lambda$.

Comparing the viscous contribution of the substrate to the elastic contribution 
of the coating defines yet another line,
\begin{equation}
n=\sqrt{\neta^{-1}}.
\label{eq:line_irrel}
\end{equation}
A case of interest would be $1\alt\neta\alt n^{-2}$, where the viscous 
contribution of the substrate is much larger than the elastic contribution of 
the coating. This should result in two plateaus, one from the viscous 
contribution of the coating at $d\approx Ql$, the other from the elastic 
contribution at $d\approx \lambda$ similar to those seen in ~\cref{fig:open_au} 
(a). However, this regime is inaccessible since $Q$ becomes inevitably large 
for small $\neta$ so that $Ql$ becomes comparable to or larger than $\lambda$ 
as discussed above. The line defined by~\cref{eq:line_irrel} is thus not shown 
in [Fig.~5] in the main text.

\section*{Generalizing [Eq. (8)] in the main text}
Even if the assumptions of $n$ being real and $\rho_0=\rho_1$ are relaxed, one 
can still gain useful insights into the scaling behavior of friction by 
constructing an equation similar to~[Eq. (8)] in the main text 
(or~\cref{eq:fric_sub} here in SM). This can be done by finding and comparing 
the bulk damping coefficients of the substrate and the coating from [Eq. (3)] 
in the main text,
\begin{equation}
\frac{\lim_{d\to0}\Gamma}{\Gamma^\infty} = n^{\prime 
3}\frac{\rho_0}{\rho_1}\left(\frac{\zela+n'\zvis  \frac{\eta_1}{\ctz'\rho_1 
l}}{\zela+\zvis \frac{\eta_0}{\ctz'\rho_0 l}}\right)
\end{equation}
with $\Re{n} = n' + \order{\omega^3} = \ctz'/\cto'+\order{\omega^3}$.

\bibliography{bibtex/miru}
\bibliographystyle{apsrev4-2}